\documentclass{ws-ijmpa}

\begin{document}

\markboth{P. Blasi, S. Gabici \& G. Brunetti}
{Gamma rays from clusters of galaxies}

%
%

\title{\bf GAMMA RAYS FROM CLUSTERS OF GALAXIES}

\author{PASQUALE BLASI}
\address{INAF/Osservatorio Astrofisico di Arcetri\\
Largo E. Fermi 5,  I50125 Firenze, Italy}

\author{STEFANO GABICI}
\address{Max-Planck-Institut f\"{u}r Kernphysik\\
Saupfercheckweg 1, 69117 Heidelberg, Germany}

\author{GIANFRANCO BRUNETTI}
\address{INAF/Istituto di Radio Astronomia \\
Via Gobetti 101, 40129 Bologna, Italy}
\maketitle

\begin{history}
\received{Day Month Year}
\revised{Day Month Year}
\end{history}

\begin{abstract}
Clusters of galaxies and the large scale filaments that connect
neighboring clusters are expected to be sites of acceleration of 
charged particles and sources of non-thermal radiation from radio 
frequencies to gamma rays. Gamma rays are particularly interesting  
targets of investigation, since they may provide precious
information on the nature and efficiency of the processes of 
acceleration and magnetic confinement of hadrons within 
clusters of galaxies. Here we review the status of viable 
scenarios that lead to the production of gamma rays from large 
scale structures and are compatible with the multifrequency
observations that are already available. We also discuss the
possibility of detection of gamma rays with space-borne telescopes
such as GLAST and ground based Cherenkov telescopes, and the physical 
information that may be gathered from such observations. 
\end{abstract}

\section{Introduction}
\label{sec:intro}
Clusters of galaxies are large collections of galaxies, gas and
especially dark matter, confined in a volume of a few $\rm Mpc^3$. 
Clusters and the filaments that connect them represent the largest 
structures in the present universe in which the gravitational force 
due to the matter overdensity overcomes the expansion of the
universe. This condition allows clusters to collapse and eventually 
go through a process of virialization at the present cosmic time. 

Rich clusters have typical total masses of the order of 
$10^{15} M_{\odot}$, mostly in the form of dark matter, while
$\sim 1\%$ is in the form of galaxies and $\sim 5\%$ is 
in the form of a hot ($T \sim 10^8 K$), tenuous ($n_{gas} \sim
10^{-3}-10^{-4} cm^{-3}$), X-ray emitting gas \cite{sarazinbook}. In
terms of energy density, the gas in the cluster is typically heated to
roughly the virial temperature, but there is room to accomodate, in
principle, a non negligible amount of non-thermal energy, in the form
of accelerated particles.  
 
A non-thermal component is in fact observed from several clusters of
galaxies, mainly in the form of a tenuous radio emission
\cite{luiginarecent}. In some cases a hard (non-thermal) X-ray
emission is also observed \cite{HXRrecent}, while more debate exists
on the non-thermal origin of a soft X-ray and UV emission \cite{softXUV,softXUV2}. 
The radio emission has a steep spectrum and appears mainly in two
forms, halos and relics, the former being very extended, fairly
regular and at the cluster center, and the latter being typically
elongated and located in the outskirts of clusters.

Roughly $20\%$ of clusters with luminosity in the
0.1--2.4 keV band $>5 \times 10^{44}$ erg/s are
observed to have radio halos \cite{halostatistic}. In general radio
halos are rather rare phenomena \cite{giovanninitordi}, which means
that their typical lifetime is relatively short ($\sim 1$ Gyr)
\cite{kuo}. Such radio emission can only be
interpreted as the result of synchrotron emission in the intracluster
medium, thereby confirming that 1) relativistic electrons are present,
2) the intracluster medium (ICM) is magnetized and 3) the magnetic field 
structure is topologically complex. The third point is inferred from 
the fact that the synchrotron emitting particles should be
diffusively trapped on the scale of radio halos \cite{petrosian01}. 

Independent evidence for turbulent magnetic fields in the ICM 
comes from Faraday rotation measurements (RM) from radio sources
located inside or behind clusters \cite{Bfield,Bfield2}.
This method leads to values of the field of the order of several
$\mu G$ \cite{clarke}, but the results are affected by uncertainties
in the topology of the 
field and the spatial distribution of the electron gas, as well as by
the subtraction of the intrinsic RM at the source. 
Smaller fields, of the order of a few tenths of $\mu G$, are obtained 
from the combination of radio and hard X--ray measurements, in the few cases
in which data at these wavelengths are available from the same cluster.
This method relies on the assumption that the 
diffuse radio emission and the hard X--ray excess are cospatial and 
produced by the same population of relativistic electrons via 
synchrotron emission and inverse Compton scattering respectively
\cite{BIC}. Although it is generally accepted that at least a fraction
of the ICM is magnetized at $\mu G$ level, the intrinsic uncertainties
in both these methods do not allow yet a precise determination of the  
magnetic field and its spatial profile \cite{rephaelifield}.

The electrons responsible for the radio emission may be accelerated in  
a variety of ways and in a variety of sources within the ICM. 
However, it is customary to classify the models of the origin
of the non-thermal activity in {\it primary electron models}, 
{\it secondary electron models} and {\it reacceleration models},
depending on whether the radiating electrons are 1) accelerated in
specific sites (e.g. shocks, AGNs, galaxies), 2) produced as secondary 
products of hadronic interactions, or 3) continuously reaccelerated
from a pre-existing population of non-thermal seeds in the ICM.  

Shock waves associated to the process of structure formation are
expected to be the sites where most cosmic rays in clusters are
accelerated. In particular, the large, strong shocks that bound the
motion of matter in filaments and during accretion onto clusters are
believed to be very efficient accelerators. 

To date, no non-thermal activity has reliably been observed from
accretion shocks or large scale filaments \footnote{Recently, a
  detection of radio waves from a cosmic shock surrounding the cluster
  A3376 has been claimed in \cite{bagchi}. However, alternative
  explanations, for instance associated with outgoing merger shock
  fronts may be equivalently consistent with the data.}. As we
discuss later, these structures may however be detected in the near
future or lead to phenomena that may be observable from within the
cluster volume.  

No cluster of galaxies has been firmly detected in gamma rays so far
\cite{olaf}. Despite this non detection, two ideas contribute to
reinforce the belief that clusters of galaxies can be interesting as
high energy radiation emitters: in \cite{bbp} and \cite{vab} it was
first understood that the bulk of cosmic rays accelerated within the
cluster volume would be confined there for cosmological times,
thereby enhancing the possibility of inelastic proton--proton
collisions and consequent gamma ray production through the decay of 
neutral pions. 

In \cite{lw} it was recognized that large scale shocks associated with
the formation of structures in the universe, may accelerate electrons
to TeV energies, implying that high energy emission would occur due to 
the upscattering of the photons of the cosmic microwave background to
gamma ray energies, through inverse Compton scattering (ICS). 
The detection of gamma rays from clusters of galaxies, or more in
general from large scale structures would represent an important and
crucial step forward towards a better understanding of the processes
of acceleration of particles in the ICM and cosmic ray confinement.

The measurement of the gamma ray flux associated with clusters can
allow us to discriminate among vastly different scenarios for the
origin of the non-thermal activity associated with the formation of
large scale structures in the universe. 

This review is organized as follows: in \S \ref{sec:confi} we
discuss the important phenomenon of cosmic ray confinement in the
ICM and the main energy loss processes for leptons and hadrons. In \S
\ref{sec:acceleration} we summarize our understanding of the sources
and acceleration sites for cosmic rays in clusters of galaxies. \S
\ref{sec:shocks} is devoted to a discussion of the formation of large
scale shocks during mergers or accretion and to assessing their role as
cosmic ray accelerators. In \S \ref{sec:fermi} we discuss some recent
developments in the theory of particle acceleration at shock waves,
and their possible role for large scale structure formation shocks. 
In \S \ref{sec:gamma} we summarize the main predictions on the fluxes
of gamma rays from clusters of galaxies, both in the GeV and TeV
energy domains. The contribution of clusters
of galaxies to the diffuse gamma ray background is discussed in
\S \ref{sec:back}. Our conclusions are provided in \S
\ref{sec:concl}.

\section{Dynamics of accelerated particles in clusters of galaxies} 
\label{sec:confi}

The propagation of charged particles (nuclei and electrons) injected
in the ICM is mainly determined by diffusion, convection and energy
losses.  

As long as the turbulent part of the magnetic field remains smaller
than the large scale field, the random walk in pitch angle that the
particles perform, and the spatial diffusion that follows are
described by the so-called {\it quasi-linear theory}
\cite{wentzel}. The diffusion coefficient for scattering of particles
with Alfv\'en waves is therefore  
\begin{equation}
D(p) = \frac{r_L (p) v(p)}{3 {\cal F}(k(p))}=
\frac{1}{3}r_L c \frac{B^2}{\int_{2\pi/r_L}^\infty dk P(k)}, 
\end{equation}
where $p$ is the particle momentum, $r_L(p)=pc/ZqB$ is the Larmor
radius for a particle of charge $Z$ in a magnetic field $B$, $v(p)\sim
c$ is the particle velocity and ${\cal F}(k(p))$ is the energy density
in the form of turbulent field on a scale $k$ that interacts
resonantly with particles with momentum $p\propto 1/k$. We introduced
the power spectrum of the perturbations of the magnetic field $P(k)$,
such that $\int_{k_{min}}^\infty dk P(k) = \xi B^2$ and $\xi\leq 1$.
The minimum wavenumber $k_{min}=2\pi/L_{max}$ is related to the
largest scale in the turbulent magnetic field. In clusters it is
usually believed that $L_{max}\sim 100$ kpc. For a  
Kolmogorov spectrum $P(k)\propto k^{-5/3}$, the diffusion coefficient
reads
\begin{equation}
D(E) \approx 7\times 10^{29} E(GeV)^{1/3} \xi^{-1} B_\mu^{-1/3}
L_{100}^{2/3}~\rm cm^2 s^{-1},
\end{equation}
where $B_\mu$ is the magnetic field in units of $\mu G$ and $L_{100}=
\left(\frac{L_{max}}{\rm 100 kpc}\right)$. 
Faraday rotation measures (RM) support a scenario in which most 
of the magnetic field energy is associated with a turbulent field
tangled on both small and large scales \cite{govoni06}.
At small scales (few kpc) a Bayesian analysis of the RM 
suggests that the power spectrum of the field follows a Kolmogorov 
scaling \cite{vogt+enss05}, although the situation at larger scales
might be more complex \cite{murgia04}. In the absence of better
estimates or measurements of the intracluster magnetic field strength
and topology, here we assume $\xi\sim 1$. Larger values of $\xi$ are
hardly treatable within quasi-linear theory but qualitatively they
lead to more effective confinement, although it is not even clear
whether the particle propagation remains purely diffusive in such a
limit. 

The diffusion time on the scale $R\sim 1$ Mpc of the whole cluster is
$\tau_{diff}=R^2/4D(E)$, which exceeds the age of the cluster 
($t_0 \approx 10$ Gyr) for energies $E \leq E_c = 1 ~ B_\mu
L_{100}^{-2} ~ \rm TeV $. Particles with energies lower than $E_c$ are
confined within the cluster volume, a phenomenon which is unique for
clusters of galaxies and is mainly due to their very large physical
size \cite{bbp,vab}.  

The prediction of diffusive confinement of cosmic rays within the
cluster volume is very strongly dependent upon the choice of the
diffusion coefficient and even the specific assumptions on the
parameters involved, due to both the energy content in the form of
turbulent magnetic field (the parameter $\xi$) and the spectrum
of fluctuations in the B-field. For instance a spectrum with
$P(k)\propto k^{-3/2}$ would lead to $E_c = 4.7\times 10^5 \xi 
~ B_\mu L_{100}^{-1} ~ \rm GeV $. The dependence on the coherence
scale of the field is also important, the confinement being more 
efficient for smaller values of $L_{max}$.

\begin{figure}[thb]
 \begin{center}
  \mbox{\epsfig{file=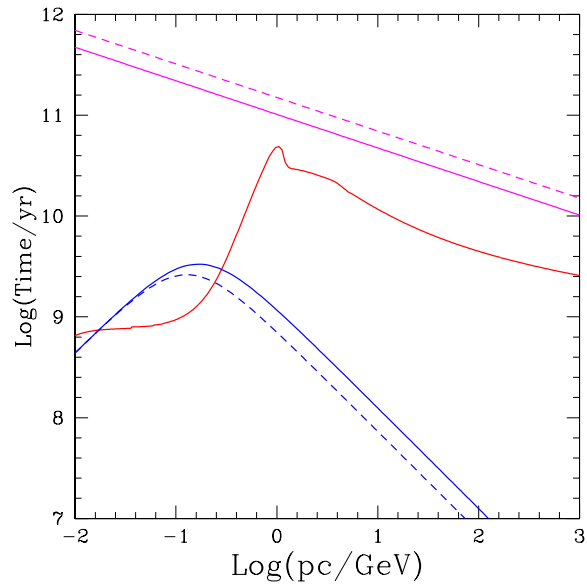,width=8.cm}}
  \caption{Time scale for energy losses of protons (middle) and 
electrons (bottom). The lines in the top part are the diffusion time
scales for protons for a Kolmogorov spectrum of magnetic
fluctuations. The magnetic field is $1 \mu G$ (solid) and $3 \mu G$
(dashed line).
The feature at 1--3 GeV in the time scale for energy losses
of protons is due to the combination of the shapes of the inclusive pp
cross section for $\pi^+$ and $\pi^-$.}   
\label{fig:losses}
 \end{center}
\end{figure}

While diffusing in the intracluster magnetic fields, charged particles
are subject to energy losses: for electrons the dominant loss
channels are ICS off the photons of the
cosmic microwave background (CMB), synchrotron emission, and Coulomb
losses (at low energies). Protons on the other hand lose energy mainly
through $pp$ inelastic scattering when their kinetic energy is higher
than $\sim 300$ MeV. At lower energies Coulomb losses become
important. The time scales for losses due to the combination of 
these processes are illustrated in Fig. \ref{fig:losses}. The curve 
in the middle is the time scale of losses for protons, while the 
bottom curves represent the loss time scale for electrons in a mean 
magnetic field $B= 1\mu G$ (solid line) and $3\mu G$ (dashed line). In  
the same plot we also show the confinement time for the case of a 
Kolmogorov spectrum with $\xi=1$ and $L_{max}=100$ kpc and mean magnetic 
field $B= 1\mu G$ (solid line) and $3\mu G$ (dashed line). 

One consequence of the confinement of the hadronic component in the
ICM is that the grammage traversed by cosmic rays is
increased to the level that a substantial gamma ray production can in
principle take place. The grammage for a cosmic ray proton with energy 
$E$ in a cluster is $X(E)= X_c = m_p n_{gas} c t_0\approx 1-10~\rm
g~cm^{-2}$ for $E<E_c$ (here $m_p$ is the proton mass and $n_{gas}$ is
the gas density). For $E>E_c$ the grammage decreases with energy. The
value of $X_c$ should be compared with the nuclear grammage 
$X_{nuc}\approx 50 ~\rm g~cm^{-2}$, which implies that an appreciable 
amount of the energy of a low energy cosmic ray is lost in inelastic 
$pp$ scatterings, leading to gamma ray and neutrino production
\cite{bbp,vab} and to the generation of a secondary population of
electrons and positrons.   

From Fig. \ref{fig:losses} we can also conclude that electrons
accelerated at specific sites can hardly radiate their energy far from
the production site \cite{jaffe}. 
This results in the fact that the region of radio
emission is very localized, quite unlike what is observed in radio
halos.  
Radio emission at $\sim 1$ GHz is typically generated by electrons 
with energy $\approx 10 B_\mu^{-1/2}$ GeV, and at this energy the
loss time is $\approx 10^8$ years.
The distance covered by diffusion during the loss time is tiny compared 
with the observed size of the radio emitting region (several hundred kpc at 
this frequency). The electrons could move farther from the production 
site due to advection (for instance behind a shock), but this would 
need that fast shocks, with Mach number larger than 4--5, should cross
the cluster cores, and still the radio emission would decay after 
$\lesssim 100$ million years, which would make radio halos less 
frequent than observed \cite{advection04}. These arguments lead 
to the requirement that electrons are generated or accelerated everywhere in
the cluster, which leaves us with two possibilities: either the
radiating electrons have a secondary origin, being produced by $pp$
scatterings \cite{dennison,blasicola}, or they are 
continuously reaccelerated {\it in situ} through a second order 
Fermi mechanism \cite{schlickeiser,tribble93,gianfranco1,petrosian01}. 
Below we discuss these possibilities in some more detail.

\section{Sources of cosmic rays and models of particle acceleration}
\label{sec:acceleration}

Whether the radiating electrons are accelerated at specific sites,
reaccelerated from pre-existing seeds or produced as secondary
products of hadronic interactions, a class of sources of accelerated
particles (electrons, hadrons or both) must exist in the ICM.
Particle acceleration can take place in several places in the
ICM, from ordinary galaxies, which contribute a total luminosity 
in the form of protons $\sim 3\times 10^{42} \rm erg~s^{-1}$, to 
active galaxies, which may plausibly contribute $\sim 10^{45}\rm
erg~s^{-1}$ over periods of $\sim 10^8$ years 
\cite{vab,colablasi,blasicola,volkatoyan99,ensslin97}. 

The process of structure formation contributes the most to the
energetics of clusters, both in the thermal \cite{cenostriker} and
non-thermal components  
\cite{bbp,vab,miniaprotons,pasqualemerger,fujitasarazin,takizawanaito,sarazinmerger,noi1,ryu,rossella}.   
Mergers between two or more clusters are observed to heat 
clusters \cite{MV01,bow} and possibly accelerate particles
\cite{roettiger,m520,c+e06}, while it is harder to measure 
the effect of the continuous infall of material onto clusters from the
colder outskirts. The two processes are not really independent, but
for simplicity we will refer to the former as {\it mergers} and to the
latter as {\it accretion}. Mergers provide the largest contribution to
the heating of the ICM, mainly through the production
of weak shocks that transform an appreciable fraction of the relative 
kinetic energy of the merging clusters into thermal energy. Accretion
is less important in this respect but it is likely to play an
important role for non-thermal phenomena (see next section). 

The total energy in a merger of two clusters with roughly equal mass
$M=10^{14}M_\odot$ is $E\approx 10^{63}$ erg. A few of these events
may lead the resulting cluster to reach the virial temperature or
close to it. It is important to realize that the majority of the mass
relevant for this estimate is in the form of dark matter, while only
baryons (and electrons) are heated. In the assumption that
the gaseous components of the initial clusters are at the virial 
temperature, it is easy to show that the merger occurs at
supersonic relative speed, therefore implying the formation of 
shock waves. The same conclusion can be reached for the accreting
material \cite{bert}. If a fraction of this energy can be converted at
the shocks into non-thermal particles through a first order Fermi
process, then the ICM may be populated with a large 
amount of non-thermal particles, which may potentially become 
important even for the dynamics of the cluster \cite{ryu}. 
Both electrons and protons can be accelerated at shocks, and secondary
electrons can in turn be produced during inelastic collisions of
cosmic ray protons in the ICM. 

In terms of the morphology of the resulting non-thermal emission these
scenarios of injection of relativistic electrons are not equivalent.
If electrons are accelerated at merger shocks, the resulting radio
emission is found to have a filamentary structure tracing the 
position of the shock \cite{miniatielectrons},
rather than the regular, symmetric structure observed in radio halos. 
That is because, as discussed in Sec. \ref{sec:confi}, electrons
cannot diffuse far away from their acceleration site, due to the short
cooling time. Radio relics are indeed interpreted as radio
emission triggered by merger shocks
\cite{roettiger,e+gk01,hrelics04}. 

Extended and fairly regular radio emission is predicted when the
radiating electrons are of secondary origin, since parent 
protons can diffuse on large scales 
\cite{dennison,blasicola,dolagensslin,miniatielectrons}. However two 
aspects of radio halos are difficult to explain
in this scenario: 1) since all clusters have suffered mergers
(hierarchical scenario) and hadrons are mostly confined within
clusters, extended radio emission should be basically observed in
almost all clusters. Radio halos are presently detected only in a
fraction of massive and merging clusters 
\cite{halostatistic,giovanninitordi,halomerger,halomerger2,luiginarecent,giaci}. 
2) Some of the observed radio halos are found to have a synchrotron 
spectrum with a cutoff at a few GHz
\cite{thierbach,giovannini93,feretti04}, corresponding to a cutoff in 
the electron spectrum.
This, together with the fact that radio halos are very extended,
would require artificial and unrealistic assumptions in order to 
be consistent with secondary models \cite{advection04,blasikorea}.
On the other hand the observational facts illustrated above are
mainly based on a handful of clusters and might reveal themselves as
incomplete when a more significant sample will become available. 
It is possible that some halos, possibly not yet identified, may
have secondary origin, despite the fact that the ones that we are
now aware of can hardly be exlained in this way.

Gamma ray emission from clusters of galaxies is still believed to
be the most effective way to detect the presence and confinement of
cosmic ray hadrons in the intracluster medium.

Present observations of extended radio emission and, when
present, of hard X-ray emission, are best explained by models of
continuous reacceleration of electrons due to particle-wave
interactions. A recent review can be found in
\cite{advection04}. Here we limit ourselves to provide a summary of
the basic aspects of this class of models.

Reacceleration models are basically models of second order Fermi
acceleration \cite{fermi,fermi2}, in which charged particles are accelerated
stochastically due to the random interaction of the particles with
perturbations (waves) in the structure of the magnetic field. These
wave-particle interactions typically require that mildly non-thermal
particles (lorentz factors $\gamma \sim 100-500$) are already present 
in the region (seed particles). In the case of clusters of galaxies 
these seeds could be provided by the past activity of active galaxies 
in the ICM or more plausibly by the past merger history of the cluster
\cite{gianfranco1}. Alternatively the seeds could be secondary
products of hadronic interactions \cite{hybrid}. 

As stressed above, the extended radio emission from clusters of 
galaxies is a rather rare phenomenon, hinting to the fact that the
episodes of radio activity are relatively short ($\sim 1$ Gyr). In the
case of wave-particle interactions, this scale is automatically
recovered as a combination of the time necessary for the cascading of
turbulence from large scales (300--500 kpc) to the small scales
relevant for particle acceleration, and of the cluster--cluster
crossing time. 

Most scenarios of particle reacceleration in the ICM
are based on particle interactions with either large scale compressible 
modes (magnetosonic) or small scale Alfv\'en modes.
Resonant particle acceleration due to Alfv\'en modes is one of the
most common mechanisms used in particle acceleration in astrophysics,
as the resonant coupling with fast particles is the main channel 
of energy dissipation of these modes. Alfv\'enic reacceleration has
been used to explain the origin of the emitting electrons in radio halos 
\cite{ohnoturb,fujitaturb}.
Detailed time dependent calculations \cite{noi4,hybrid} suggest that
efficient Alfv\'enic reacceleration of relativistic electrons might
operate in the ICM provided that the energy budget of relativistic
hadrons is below 3-5\% of that of the thermal ICM and that a few
percent of this thermal energy is channelled into the modes during a
lifetime of a radio halo.  
Relativistic hadrons affect the reacceleration of fast electrons 
because they provide the dominant contribution to
the damping of Alfv\'en modes in the ICM: the acceleration of such
hadrons increases the damping of the modes and this limits 
the acceleration of relativistic electrons
({\it wave--proton boiler}) \cite{noi4}.
The injection process of Alfv\'en modes in the ICM and the assumption
of isotropy of these modes represent the most important source of 
uncertainties in this class of models. Alfv\'en waves couple with
relativistic electrons (and protons) on very small spatial scales 
$\l \sim 2\pi p/(\Omega m)$ ($p$, $m$ and $\Omega$ being the momentum, 
mass and Larmor frequency of particles) and this requires a surprisingly 
efficient Alfv\'enic cascade if these modes are connected with 
the merger--driven turbulence on larger scales.
In addition, the cascading of MHD turbulence in nature usually
develops into an anisotropic spectrum on scales significantly smaller
than the injection scale \cite{goldreich,cho} and this should strongly
reduce the efficiency of particle acceleration \cite{yan}.
On the other hand, injection of Alfv\'en modes at (quasi--) resonant
scales might occur as a consequence of the cascading of fluid
turbulence \cite{kato}, or of several kinds of plasma instabilities
\cite{beresnyak}; the physical details and efficiency of these
mechanisms are however yet to be understood.

Some of the problems encountered with Alfv\'en modes may be solved 
if the interaction of particles with compressible modes is considered, 
although new issues are also raised. Compressible modes do not require 
injection on small spatial scales: 
a substantial fraction of the cluster turbulence is expected to be
in the form of large scale compressible isotropic modes, which 
might efficiently accelerate fast particles via resonant Transit Time
Damping (TTD) and non--resonant turbulent--compression
\cite{rossella,cho2,gianfrancoprep}. 
The main source of uncertainty in these models is our ignorance of
the viscosity in the ICM. Such viscosity could in fact severely damp
compressible modes on large scales, inhibiting particle acceleration.
On the other hand, turbulence in the ICM is subsonic but strongly 
super-Alfv\'enic, and in these conditions the bending of the magnetic 
field lines is expected to maintain the effective ion--ion mean free 
path much smaller than that in the classical unmagnetized case 
\cite{lazarian06}. This phenomenon should limit the viscous dissipation,
so that collisionless dampings with the thermal and relativistic 
particles should represent the main source of turbulent dissipation.
In this case detailed calculations of resonant--TTD and non--resonant
stochastic coupling with fast modes in the ICM suggest that 
the resulting efficiency of the particles re--acceleration 
could be sufficient to power radio halos, provided that 
the rms velocity of these compressible eddies is 
$V_{rms}^2 \sim (0.1-0.3) c_s^2$ \cite{gianfrancoprep}.
Here, at variance with the Alfv\'enic case, the acceleration
efficiency of electrons and positrons is not affected by the presence
of relativistic hadrons since the damping of compressible modes
in the ICM is largely dominated by the TTD--coupling with thermal
electrons and protons \cite{gianfranco06,gianfrancoprep,rossella}. 

In models with stochastic reacceleration, the spectrum of electrons
has typically a cutoff at Lorentz factors $\leq 10^5$, therefore no
appreciable ICS gamma rays may be expected. However, as pointed out in 
\cite{hybrid}, the role of protons in these models may be very
important and one may reasonably expect gamma ray emission from either
reaccelerated protons or through ICS of the secondary electrons
generated in inelastic $pp$ collisions. Moreover, stochastic 
reacceleration may significantly modify the shape of the spectrum of 
relativistic hadrons \cite{noi4,gianfrancoprep}, thereby changing 
the shape of the spectrum of neutral pions and gamma rays. Further 
investigations of these effects are certainly needed and desirable.

\section{Large scale shocks during cluster formation}
\label{sec:shocks}

As discussed above, most of the non-thermal energy content 
of clusters is believed to be associated with the process of large
scale structure formation. We briefly review this process here, with 
particular emphasis on the issues which are most important for
particle acceleration and related non-thermal phenomena in clusters.  

A wide consensus has been reached in the community on the hierarchical
model of structure formation, which predicts that increasingly larger
halos are formed by merging of smaller halos. On average smaller
structures are therefore formed at earlier times than larger
structures. 
During mergers between clusters, shocks are driven by gravity in the
diffuse baryonic component, which is heated up to X-ray temperatures 
\cite{sarazinbook}. High resolution X--ray observations of clusters 
of galaxies performed with Chandra and XMM have provided us with direct
evidence for disturbed X-ray morphology and hot shocks associated 
with merging clusters \cite{buotebook,formanbook}. 

While a complete understanding of the process of structure formation
can only be achieved through numerical N-body simulations, a simple
analytical description is useful in that it allows us to 
check the main numerical results versus some basic physical
expectations.   
These analytical descriptions come in different flavors and are widely
discussed in the literature. Historically, the first approach to the
problem was proposed by Press \& Schechter (hereafter PS)
\cite{PS} and further developed in \cite{bond,LC} among others. 

The PS formalism provides us with the differential comoving number
density of clusters with mass $M$ at cosmic time $t$, $n(M,t)$, and
with the rate at which clusters of mass $M$ merge at a given time $t$
to give a cluster with final mass  $M^{\prime}$, $R(M,M^{\prime},t)$. 
These two distributions can be randomly sampled to construct 
Monte Carlo realizations of the merger history of a cluster. 
This procedure is illustrated in the left panel of Fig. 2 
\cite{noi1} for a cluster with present mass of
$10^{15} M_{\odot}$. The history is followed backward in time  
to a redshift $z = 3$. The big jumps in the cluster mass correspond 
to major merger events, in which two subclusters collide to form 
a bigger object, while smaller jumps correspond to accretion events 
in which the cluster mass is increased by a small amount.  

The PS formalism has been extensively used in the past to investigate
the heating of the ICM and other thermal properties of
clusters \cite{cavaliere,randall}. 
In \cite{noi1} such approach has been generalized in order to describe
the acceleration of particles at merger shocks and to investigate the
implications for clusters non-thermal emission. 
The novelty in the approach resides in the combination of the PS
formalism with a recipe to evaluate the
Mach number of merger related shocks. An accurate determination of the
shock strength is of crucial importance for predicting the non-thermal
behavior of clusters of galaxies, since the spectrum of the particles
accelerated at a shock is, at least in the test particle
approximation, uniquely determined by its Mach number. 

The Mach number of merger--related shocks can be estimated by using an
approach introduced in its original version in \cite{takizawa} and 
\cite{noi1}. The basic assumption is that the two merging clusters,
with masses $M_1$ and $M_2$ are virialized to start with. 
The virial radius for each cluster can be written as follows: 
\begin{equation}
r_{vir,i} = \left(\frac{G M_i}{100 \Omega_m H_0^2
  (1+z_{f,i})^3}\right)^{\frac{1}{3}} \sim 3 ~ \left(
\frac{M_1}{10^{15}M_{\odot}} \right)^{1/3} (1+z_f)^{-1} Mpc, 
\label{eq:vir}
\end{equation}
where $i=1,2$, $H_0$ is the Hubble constant and $\Omega_m$ is the 
matter density fraction. The cluster formation redshift $z_f$ is on
average a decreasing function of the mass, meaning that smaller
clusters are formed at larger redshifts, although fluctuations in the
value of $z_f$ from cluster to cluster at given mass are present due
to the stochastic nature of the merger tree. 

The relative velocity of the two merging clusters, $V_r$, can be easily 
calculated from energy conservation:
\begin{equation}
-\frac{G M_1 M_2}{r_{vir,1}+r_{vir,2}} + \frac{1}{2} M_r V_r^2 = 
-\frac{G M_1 M_2}{2 R_{12}},
\label{eq:Vrel}
\end{equation}
where $M_r=M_1 M_2 / (M_1+M_2)$ is the reduced mass and $R_{12}$ is the 
turnaround radius of the system, where the two subclusters are
supposed to have zero relative velocity. In fact the final value of
the relative velocity at the merger is quite insensitive to the exact
initial condition of the two subclusters, being quite close to the
free fall velocity.  

The sound speed in the cluster $i$ is given by \cite{takizawa}
$$
c_{s,i}^2 = \gamma_g (\gamma_g - 1) \frac{G M_i}{2 r_{vir,i}},
$$
where the virial theorem has been used to relate the gas temperature to the 
mass and virial radius of the cluster. The adiabatic index of the gas is
$\gamma_g=5/3$. The Mach number of each cluster while moving in the volume
of the other cluster is then ${\cal M}_i = V_r/c_{s,i}$.
Here it has been implicitely assumed that the Mach number remains
constant during the whole merger. 

The procedure illustrated above can be applied to a generic couple of 
merging clusters, and in particular it can be applied to a generic merger
event in the history of a cluster with fixed mass at the present time.

\begin{figure}[t]
 \begin{center}
  \mbox{\epsfig{file=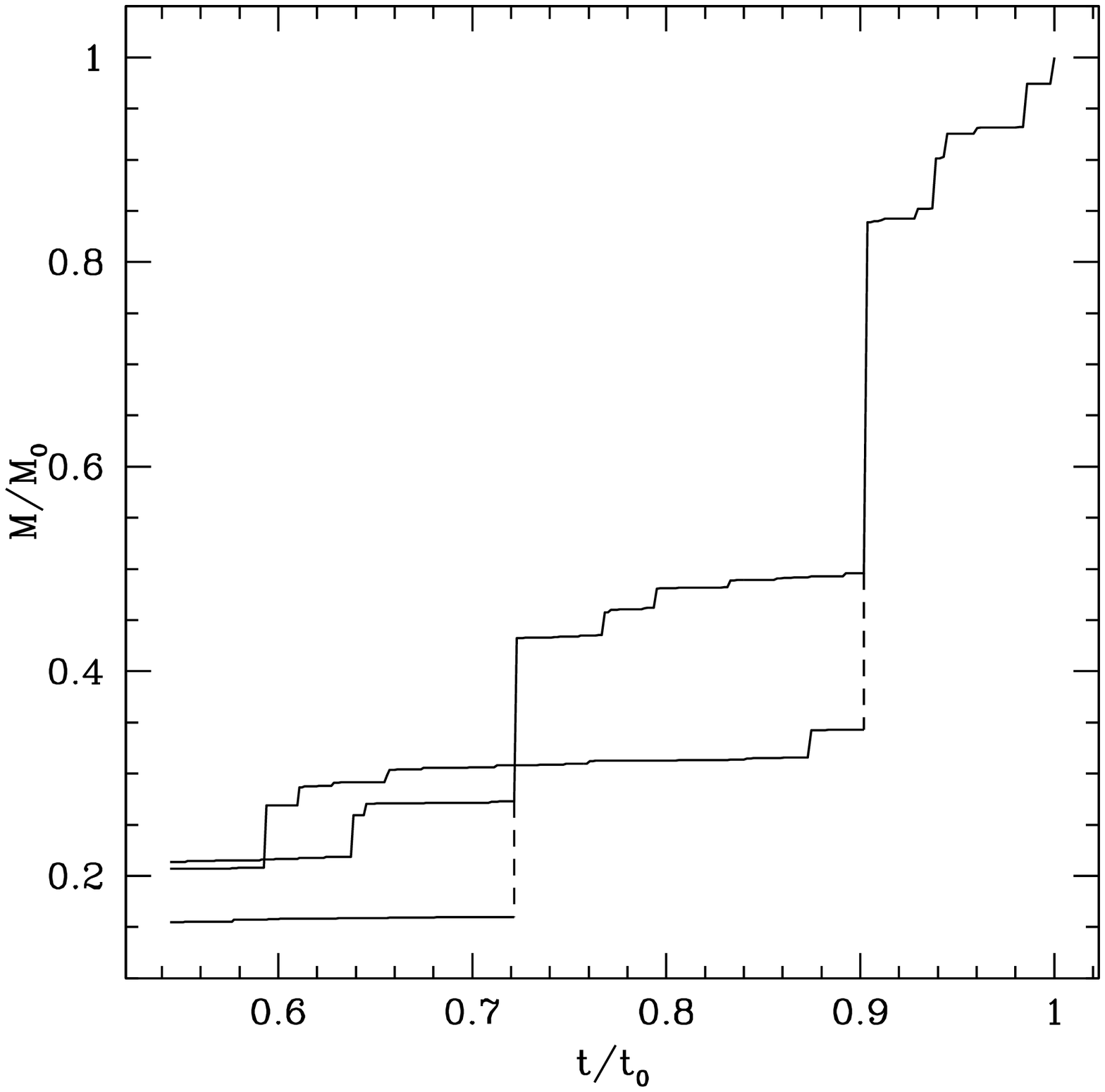,width=6.cm}}
  \hspace{0.3cm}
  \mbox{\epsfig{file=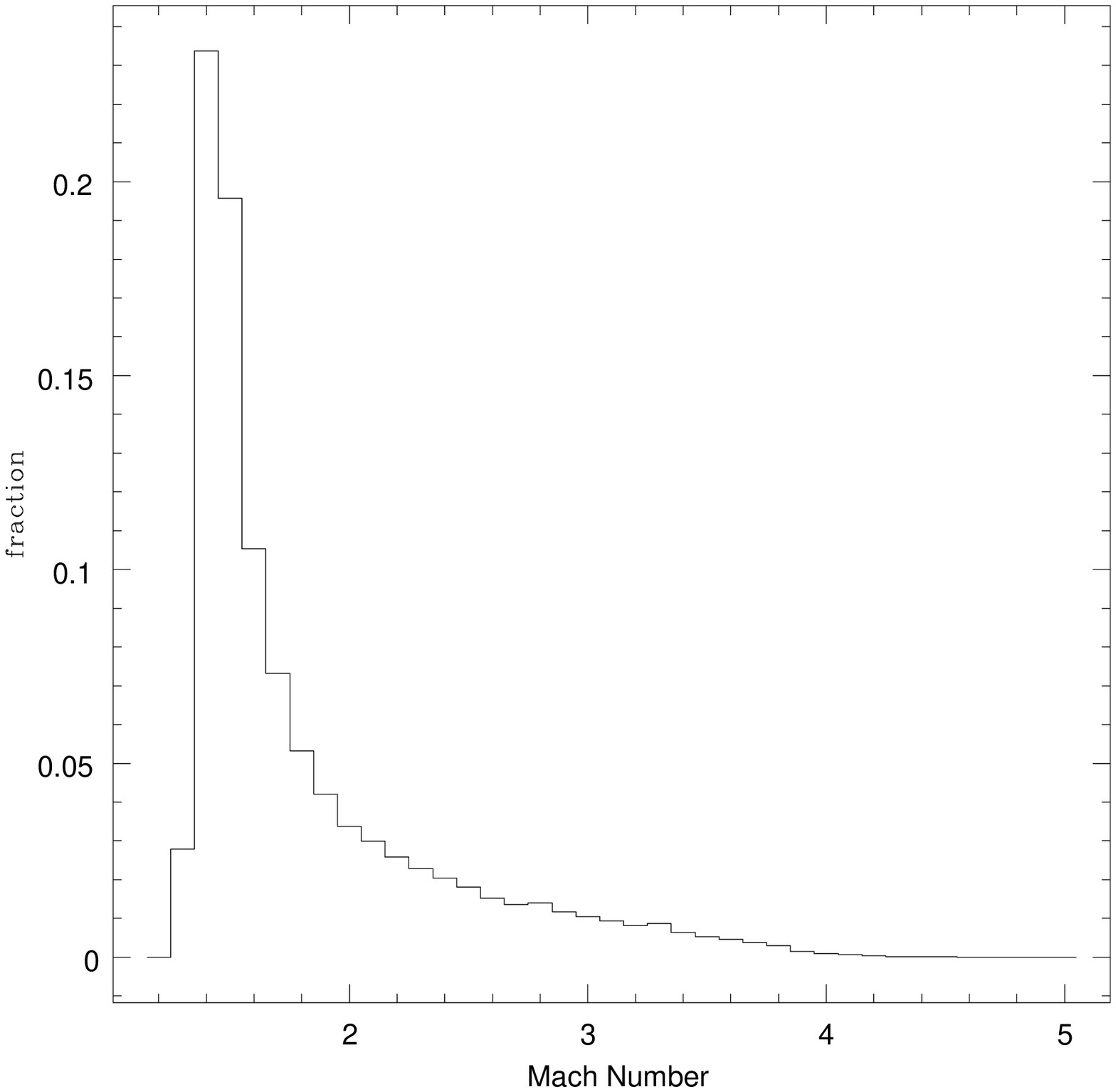,width=6.cm}}
  \caption{LEFT PANEL: Merger history of a cluster with present mass
    $10^{15}$ solar masses. The mass (y-axis) suffers major jumps in
    big merger events. Time is on the x-axis. RIGHT PANEL:
    Distribution of the Mach numbers of merger related shocks for 500
    realizations of the merger tree of a cluster with present mass
    $10^{15} M_{\odot}$.}  
 \end{center}
\label{fig:mergertree} 
\end{figure}

In Fig. 2 (right panel) the differential distribution of shock 
Mach numbers is shown. Results have been extracted from 500 
realizations of the merger history of a cluster with present mass
$10^{15} M_{\odot}$ \cite{noi1}. 
The most striking feature of the histogram is a very pronounced peak
at Mach number $\sim 1.4$. This means that merger shocks are
generally very weak, most of them are mildly supersonic, and only
$\sim 5\%$ of merger shocks populate the tail of the distribution,
with Mach number above $\sim 3$ \cite{noi3}. 

The Mach number may be higher than that provided by the procedure
illustrated above if the merger takes place in an overdense region
such as a supercluster, or if the whole temporal evolution of the
merger is considered instead of a mean value during the merger event.
However, it has been demonstrated that none of these effects alters
significantly the results illustrated above: the former effect is
negligible for very massive clusters \cite{noi1}, and 
the latter leads only to a slight increase of the merger velocity, 
less than a factor of two, but limited to a small fraction of the 
duration of the merger event \cite{berrington}. 
Some additional uncertainties might arise from the assumption that the 
merging clusters are virialized to start with and from the assumptions
made on the spatial distribution of dark matter in the parent halos. 

Due to the intrinsic complexity of the process of structure formation,
numerical simulations are the best way to study properties of
cosmological shock waves such as the shock velocity, strength and
morphology, although much care is needed in the identification of
shocks and in evaluating their efficiency as particle accelerators.
Results from earlier simulations \cite{miniatishocks} were
in clear disagreement with the above mentioned semi-analytical
findings. The distribution of Mach numbers reported in 
\cite{miniatishocks} for shocks located in the inner $0.5 h^{-1} Mpc$ 
region of clusters exhibited a pronounced peak at ${\cal M} \sim 4
\div 5$, weaker shocks being virtually absent. More recently, higher 
resolution simulations led to a revision of these earlier
results, and significantly mitigated the discrepancy between the two
approaches. According to \cite{ryu}, there is a peak in the
distribution of shock surfaces in the whole simulated
box at ${\cal M} \sim 1.5$, in very nice agreement with
Fig. 2. Moreover, the same authors showed that most
of the shock energy is dissipated at weak shocks (${\cal M} \sim 2
\div 3$). These results have been confirmed in \cite{ensslinshocks}, 
where energy is found to be mostly dissipated at shocks with Mach
number $\sim 2$.   

X-ray observations have allowed the measurement of the Mach number of 
merger-related shocks in the inner regions of a few clusters. To date, 
the strongest shock ever observed in the ICM has a Mach number of the 
order of $\sim 3$ \cite{bow}, while claims of detection of weaker
shocks have been reported \cite{shocksX,shocksX2,m520}. 

Besides mergers, the growth of structures proceeds through the 
so-called \textit{secondary infall} onto already formed objects
\cite{bert}, or {\it accretion}.
Clusters continuously attract unbound matter in their neighborhood and
their mass increases due to accretion of this new material. 
Accretion shocks form around clusters and propagate outwards,
carrying the information on the virialization of the inner regions. 
The existence of non virialized gas outside clusters is suggested by
cosmological simulations \cite{cenostriker}. 
This gas is structured in low density, relatively cold filaments and
sheets ($T \sim 10^4 \div 10^6 K$) whose X--ray emission is expected
to be tenuous and hardly detectable by present day instruments.  
The low temperature of the gas implies that accretion shocks are
always rather strong, with Mach numbers that may exceed $\sim 10$ 
\cite{ryu,pavlidou}.
The fraction of the total cluster mass accreted through these shocks
is less than $\sim 10\%$ \cite{keshet2}. This implies that, even if
accretion shocks are very strong and thus able to dissipate
efficiently the kinetic energy of the accretion flow, they do not
contribute significantly to the total energy budget of clusters
\cite{noi2,ryu,ensslinshocks}.  

\section{Diffusive particle acceleration at merger and accretion
  shocks} 
\label{sec:fermi}

A vast literature exists on the problem of diffusive particle
acceleration at collisionless shocks \cite{shockreviews1,shockreviews2,shockreviews3,shockreviews4,shockreviews5}.
In its {\it test particle} version, the theory is simple and
predicts that the spectrum of accelerated particles is a power law
$N(E)\propto E^{-\gamma}$, with a slope which is determined uniquely
by the Mach number ${\cal M}$ of the supersonic fluid,
$\gamma=2\frac{{\cal M}^2+1}{{\cal M}^2-1}$. For shocks with Mach
number ${\cal M}=2-4$ the
spectrum of the accelerated particles varies between $E^{-3.33}$ and
$E^{-2.27}$. For ${\cal M}=1.4$, where the peak of merger related
shocks was found, the spectrum is as steep as $E^{-6.2}$: such weak
shocks can only play a role as heaters of the intergalactic medium,
while being irrelevant for particle acceleration. As we discuss below,
the efficiency of injection at such weak shocks is also expected to be
negligibly small. The convolution between the efficiency as particle
accelerators and the steepeness of the spectrum selects the merger
related shocks that are most relevant for particle acceleration in the 
ICM \cite{ryu}.      

The process of acceleration of hadrons and electrons during cluster
mergers has different peculiar aspects. The time scale for energy
losses of electrons is shorter than the mean time between two
mergers. Thus, it is a good approximation to assume that
only the last merger event contributes to accelerate electrons and
that these radiate for approximately their loss time. The spectrum of
the accelerated electrons at each shock is a power law, with a slope
which is expected to be hard enough to produce appreciable non-thermal
emission only in the few ($\approx 5\%$) mergers characterized by a
relatively high Mach number (${\cal M} \gtrsim 3$) \cite{noi2}.  

The confinement of hadrons in the cluster volume for
cosmological time scales implies that any new merger related shock
front, besides accelerating a new population of particles from the
thermal pool, also re-energizes previously accelerated nuclei
\cite{noi1}. As a 
consequence, the spectrum of the hadrons present at any time in a cluster 
is the result of both acceleration and reacceleration of all
previously confined hadrons. The combination of these effects makes
the spectra of confined hadrons depart from a power law \cite{noi1},
being steeper at low energies and gradually flatter at higher
energies. This result was obtained by making the simple but most
likely unrealistic assumption that the acceleration efficiency is
independent of the Mach number (see below for a discussion of the
problem based on the non--linear theory of shock acceleration). 
It is worth stressing that while the efficiency of a
shock as a particle accelerator depends on the unknown details of
injection, the efficiency for reacceleration is known \cite{bell78}.

The situation changes dramatically if one considers strong accretion
shocks. According with the standard (test particle) theory of particle
acceleration, these shocks should result in a universal spectrum
$\propto E^{-2}$ since their Mach number is expected to be very
large. On the other hand, in these conditions test-particle theory is
known to fail, to an extent which depends upon several aspects of
the problem (maximum energy, Mach number, presence or absence of
turbulent heating, efficient generation of waves through streaming
instability). The violations of test particle theory are grouped
together in what is named {\it shock modification}, which is mainly
determined by the dynamical reaction of the accelerated particles
on the shock structure. 

The development of a non-linear theory of diffusive particle
acceleration at shock fronts, which allows to accurately describe the 
shock modification and at the same time calculate the spectrum of the 
accelerated particles, is one of the major developments in the
investigation of the process of particle acceleration in the last 
decade (see \cite{shockreviews1,shockreviews2,shockreviews3,shockreviews4,shockreviews5} for recent reviews).  
The main points of the non-linear theory of diffusive shock
acceleration can be summarized as follows: 1) strong shocks are
efficient accelerators, to the point that the dynamical reaction of
the accelerated particles cannot be neglected; 2) this dynamical
reaction is responsible for a gradual slowing down of the fluid
upstream of the shock front, the so-called {\it shock precursor},
which reflects the spatial distribution of the pressure of the
accelerated particles; 3) the precursor results in a non-power law,
concave spectrum, flatter than the test-particle prediction
($N(p)\propto p^{-4}$) at high energies and steeper than that at low
energies; 4) as a consequence of efficient particle acceleration, less
energy is available for heating of the downstream gas, which therefore
is predicted to be colder than expected on the basis of the standard
Rankine-Hugoniot relations. 

The shock modification induced by the dynamical reaction of the
accelerated particles leads to a determination of the efficiency for
particle acceleration as a function of the Mach number of the shock,
which is crucial for assessing the role of shock fronts for particle
acceleration during structure formation. 

Some of the implications of the theory for large scale structures have
been discussed in \cite{noiHD,kangjones} and are now recognized as 
crucial to understand the role of cosmic rays in clusters of galaxies
and their neighborhood. In order to illustrate the main results, we
adopt here the approach described in \cite{noiHD}, where injection
was treated following Ref. \cite{vannoni}.

\begin {figure*}[t]
\vskip -0.6cm
\centerline{\epsfysize=5.8cm\epsfbox{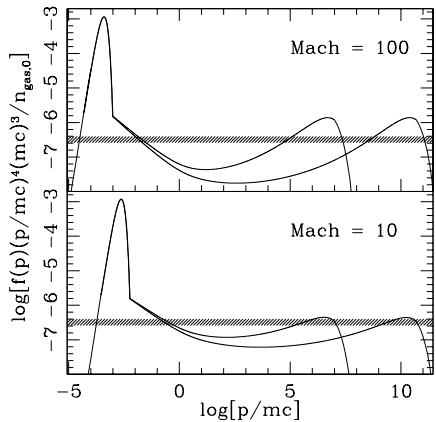}}
\label{fig:protons}
\caption{Proton spectra at the shock location for different values of 
the shock Mach number and of the maximum momentum of the accelerated 
particles.}
\vskip -0.5cm
\end{figure*}

In Fig. 3 we plot the spectra of protons at the shock location for  
Mach number $M=100$ (upper panel) and $M=10$ (lower panel) and for 
$E_{max}/(m c^2) = 10^7$ and $5\times 10^{10}$. 
These values of the Mach number are clearly suitable to describe
accretion shocks rather than merger shocks. 
The maximum energy of particles accelerated at accretion shocks is
believed to be within the range of values considered here
\cite{Emaxclusters,Emaxclusters2,Emaxclusters3}. The normalization of the curves  
is such that these spectra reproduce those expected for protons 
accelerated at the accretion shock of a Coma-like cluster and therein
confined \cite{noiHD}. The concave shape of the spectra, typical of
non-linear particle acceleration at shocks, is clear in the figure. 

The Maxwellian distribution in Fig. 3 represents the 
spectrum of the particles in the downstream fluid, at the temperature 
which is calculated by solving the suitable conservation equations at
the shock. 
As a comparison we also plot the prediction of test--particle theory
assuming that a fraction of $10\%$ of the shock kinetic energy flux is
converted into cosmic rays (horizontal lines). 
 
As stressed above, one of the consequences of the non-linear backreaction 
of cosmic rays on the shocks induced by large scale structures is that
the gas is heated less than it would be at shocks where cosmic rays are
absent. In Fig. 4 (left panel), we plot the 
temperature ratio between downstream infinity and upstream infinity
$T_2/T_1$ for an ordinary shock (dotted line) and for a cosmic ray
modified shock, for $E_{max}=10^2,~10^6$ and $5\times 10^{10}$ GeV (from 
top to bottom) as a function of the Mach number of the shock. We can
see that the temperature jump is basically unaffected by the presence
of accelerated particles for low Mach numbers, but the heating is
severely suppressed for strongly modified (high Mach number) shocks.

\begin {figure*}[t]
\vskip -0.6cm
\centerline{\epsfysize=5.8cm\epsfbox{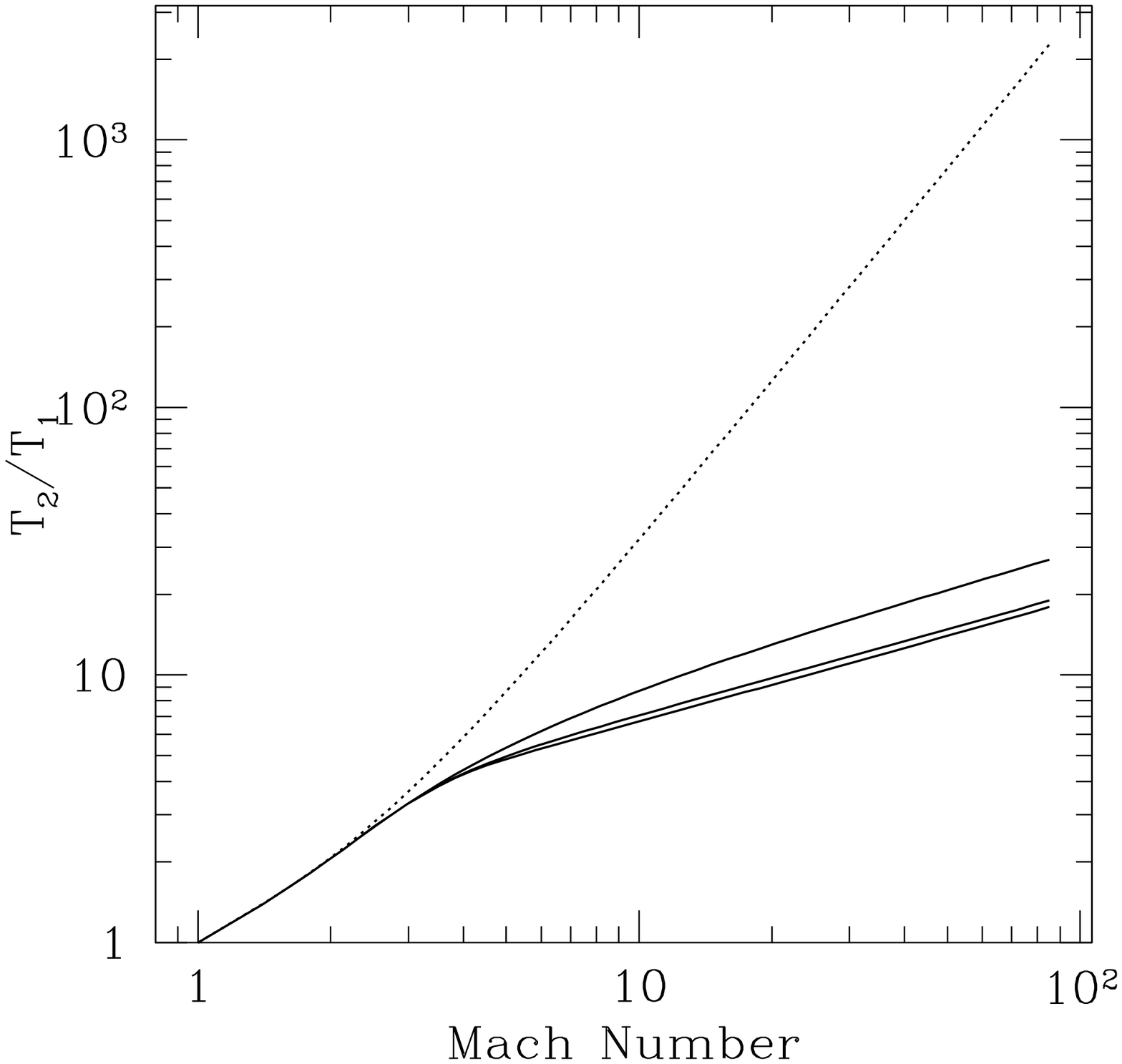}
\epsfysize=5.8cm\epsfbox{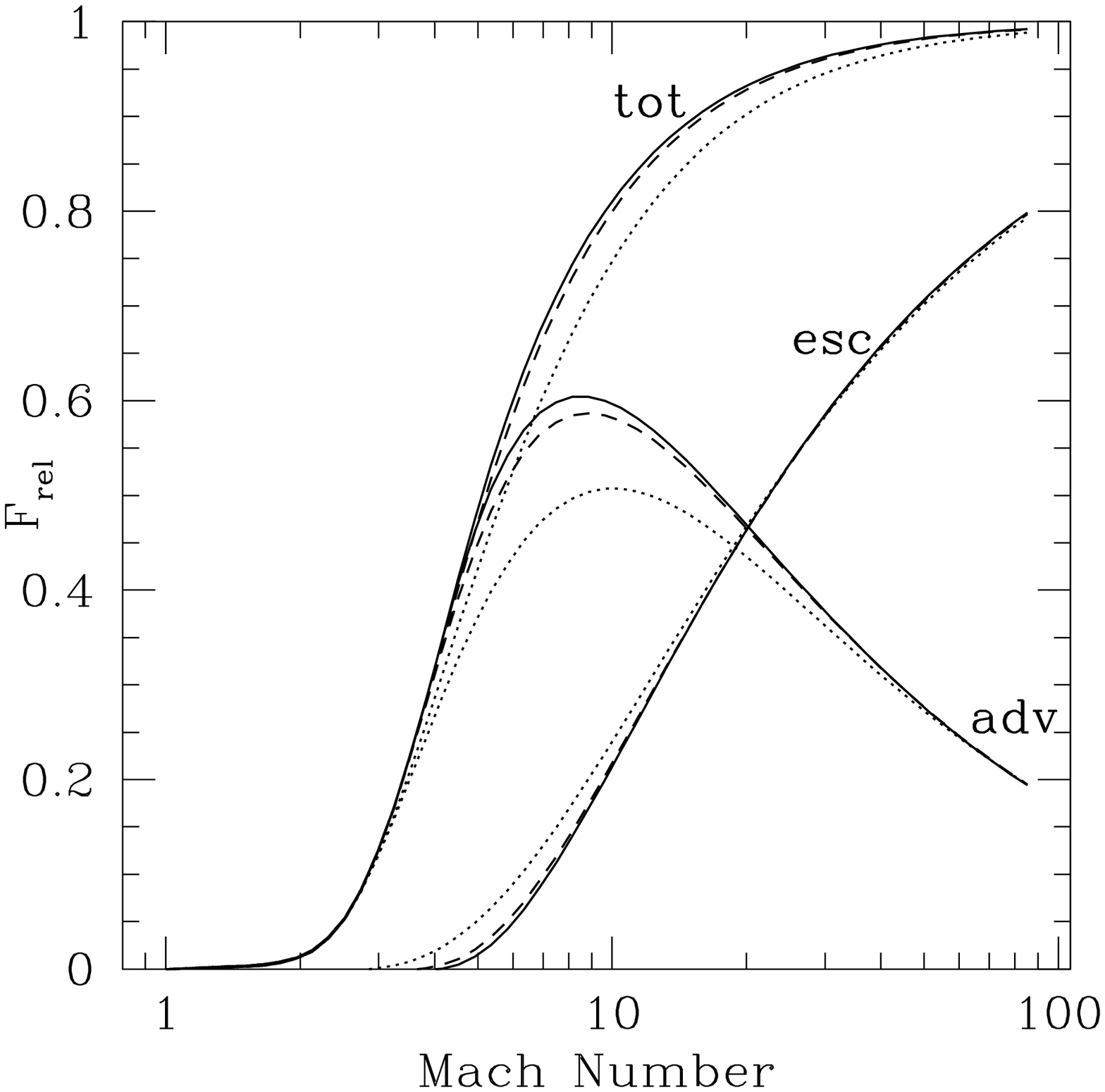}}
\label{fig:temp}
\caption{a) Temperature ratio between downstream and upstream infinity
as a function of the Mach number (see text). b) Efficiencies for advected 
particles, escaping particles and total efficiency as functions of the Mach 
number.}
\vskip -0.5cm
\end{figure*}

In Fig. 4 (right panel) we also plot the fraction of flux in the form
of relativistic particles (in units of $(1/2)\rho u^3$) which is
effectively advected downstream, the fraction of flux which escapes 
the system at $E_{max}$ (from upstream) and the sum of the two, which 
saturates to a number very close to unity for large Mach numbers. 
Solid, dashed and dotted lines refer to $E_{max} = 5 \times 10^{10}$, 
$10^6$ and $10^2$ GeV respectively. 

It is worth stressing that the very poorly known phenomenon of
turbulent heating of the upstream plasma may somewhat reduce the
shock modification with respect to the results illustated above,
though not affecting the basic conclusions.

The non-linear effects of diffusive particle acceleration are known to 
be crucial in the context of particle acceleration at supernova
remnant shocks \cite{shockreviews1,shockreviews2,shockreviews3,shockreviews4,shockreviews5}. 
In the case of clusters of galaxies these effects have never been
explored in detail, all existing calculations being based on very
simple assumptions on the spectrum of the accelerated particles (see
however \cite{noiHD}).

As discussed in \cite{noiHD}, since non--linear effects in shock
acceleration affect both the spectral shape and the overall
normalization of the accelerated particles, modifications in the
predicted gamma ray fluxes from clusters are expected
accordingly. This issue deserves further investigation, and it will
likely constitute one of the most interesting future developments in
the field.  

\section{Gamma Rays from clusters}
\label{sec:gamma}

The initial motivation for the interest in the gamma ray emission from
clusters arose as a consequence of the possibility of cosmic ray
confinement in the ICM, and consequently the possibility that 
radio halos could be the result of synchrotron emission from 
secondary electrons. A natural byproduct of the confinement is the
copius emission of gamma radiation due to the production and decay 
of neutral pions. Calculations of this emission from single clusters 
of galaxies have been carried out in 
\cite{ensslin97,colablasi,blasi99,pfrommer,atoyanvolk00,miniatisingle}. 

Gamma radiation in the ICM can also be generated as a result of 
ICS of high energy electrons off the universal photon background. 
In particular, electrons can be accelerated at cosmological shocks 
up to energies of tens of TeV. The resulting ICS emission extends 
to multi-TeV energies \cite{lw}. 

Other mechanisms of production of gamma rays in the ICM include ICS
from secondary electrons, non-thermal bremsstrahlung and ICS from
pairs generated in Bethe--Heitler processes between Ultra High Energy
protons (if present) and photons in the cosmic microwave background. 
The first two channels may be relevant only for gamma ray 
energies below $\approx 100 MeV$ \cite{pasqualemerger} and will not be
discussed further here. The latter mechanism will be briefly discussed
in Sec. \ref{sec:ultra}.    

\begin {figure*}[t]
\vskip -0.6cm
\centerline{\epsfysize=5.8cm\epsfbox{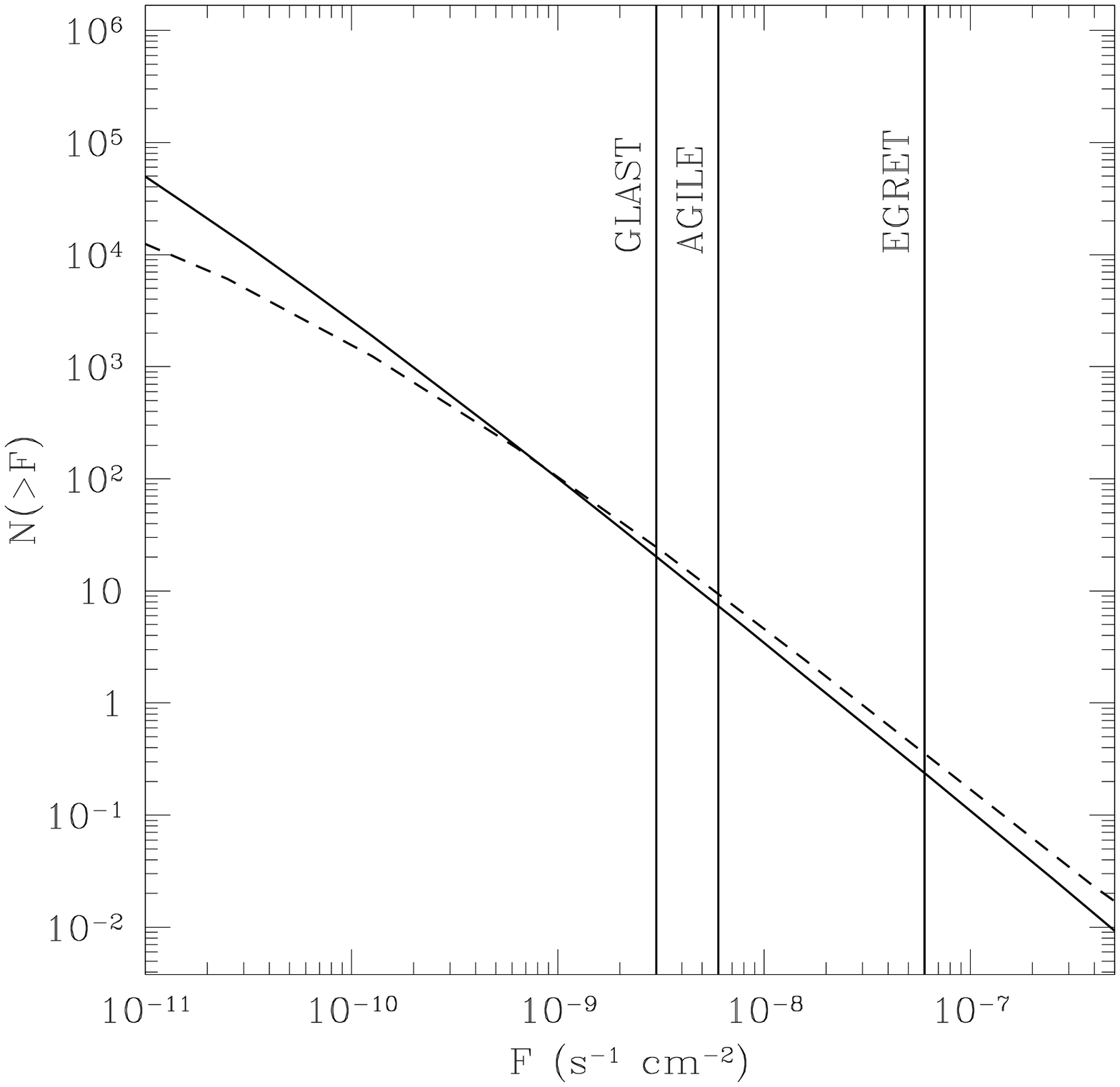}
\epsfysize=5.8cm\epsfbox{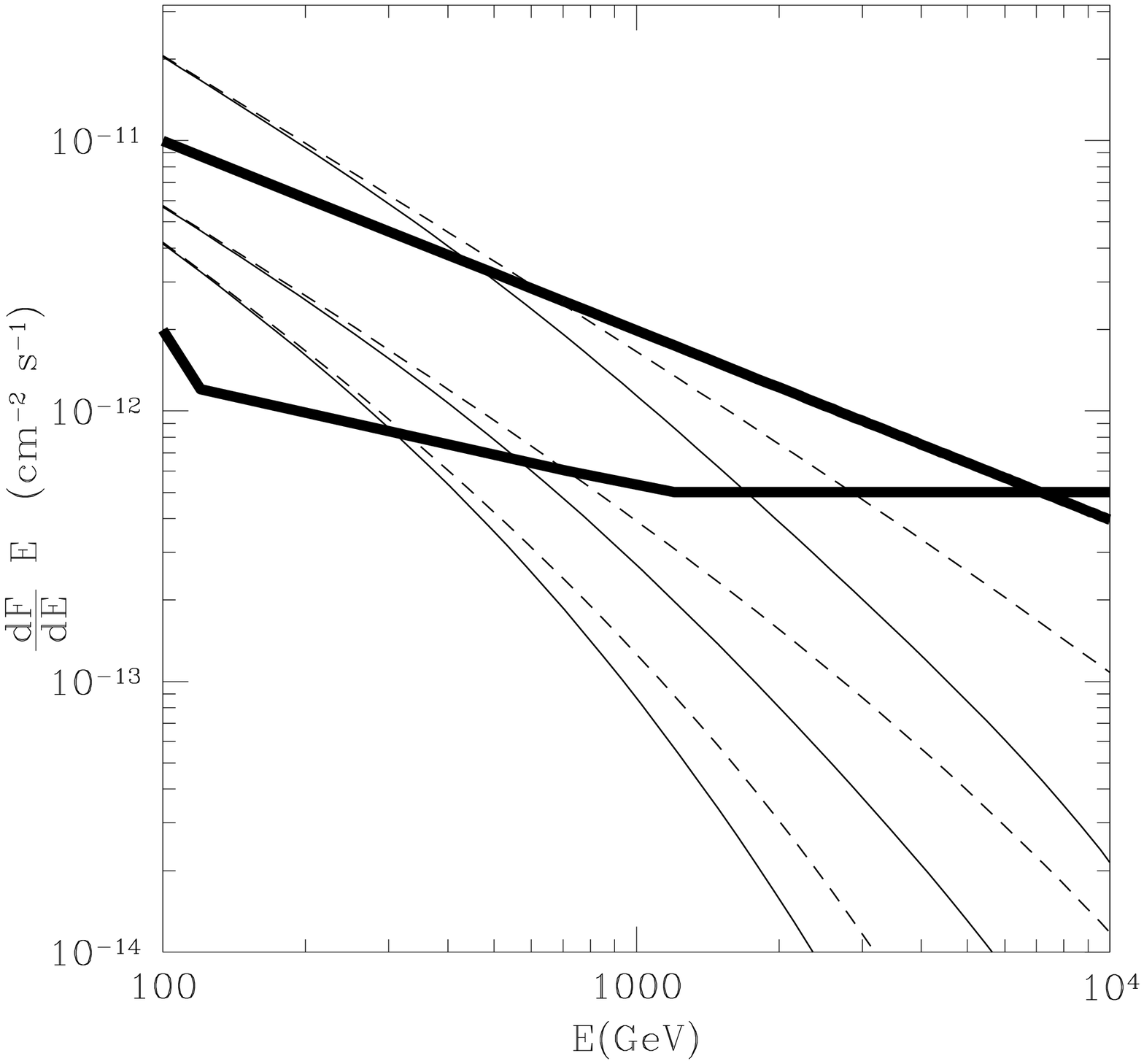}}
\label{fig:logNlogS}
\caption{a) Number of accreting (solid line) and merging (dashed line) 
clusters with gamma ray flux greater than $F$. The vertical lines 
represent the GLAST, AGILE and EGRET sensitivity for point sources. b)
Gamma ray emission in the 100 GeV - 10 TeV region. The thick solid
lines represent the sensitivities of a IACT for point sources (lower curve) 
and extended sources (upper curve). The predicted gamma ray fluxes from a 
Coma-like cluster at a distance of 100 Mpc with and without absorption of 
the infrared background are plotted as dashed and solid lines respectively.
}
\vskip -0.5cm
\end{figure*}

\subsection{Gamma rays from $\pi^0$ decay}

The present observational situation concerning radio halo emission
from clusters is not favorable for secondary electron models, at least
in those few cases in which the data are good enough to allow for such
a type of analysis, as for the Coma cluster. In such cases the radio
emission can be used to impose limits on the amount of hadronic cosmic
rays trapped in the ICM, and therefore on the flux of gamma rays due
to $\pi^0$ decay that may be expected \cite{reimers}. These limits are
usually rather model dependent, in that the amount of cosmic ray
energy that is allowed depends on magnetic field strength, slope of
the proton spectrum and spatial distribution of cosmic rays in the
intracluster medium. With these limitations, current limits suggest 
that not more than $\sim 10\div 30\%$ of the cluster thermal energy
may be in form of relativistic particles.  
If the amount of relativistic particles is not much smaller than this
limit, GLAST should easily detect the gamma rays from pion decay from
several rich clusters \cite{colablasi,reimers} in the energy range
$100$ MeV - $10$ GeV. Other limits on the gamma ray emission of some
nearby clusters of galaxies were calculated by using the existing
EGRET upper limits \cite{blasi99,pfrommer}. 

Clusters of galaxies are however interesting targets even for Imaging
Atmospheric Cherenkov Telescopes such as HESS, MAGIC and VERITAS. We
briefly consider here the perspective for detection of a Coma-like
cluster at TeV energies. 
If cosmic rays have a power law spectrum in the form $N(E) =N_0
E^{-2}$ (for the purpose of estimating the flux we are neglecting here
the non-linear effects discussed in the previous section), and are
spatially distributed following the thermal gas, the 
expected gamma ray flux is $E^2 F(E) \approx 10^{-12} (\eta/0.1)
erg/cm^2/s$, where $\eta$ is 
the fraction of the cluster thermal energy in the form of relativistic
protons \cite{atoyanvolk00,blasi99}. The spectrum of the gamma radiation is
also a power law at large enough energy, with a high energy cutoff
which may be due to either lack of confinement of the parent protons
or to absorption of gamma rays in the infrared background (in fact
  in the former case a steepening is expected rather than a cutoff). 
At the low redshift of the Coma cluster ($z = 0.023$) the opacity due
to absorption becomes of order unity at energies above several TeV
\cite{IRabsorption}.  

The expected brightness profile should be centrally peaked, tracing
the spatial distribution of the thermal gas, which represents the
target for pp collisions. Thus, the emission is expected to be
concentrated within a core of apparent size of $\sim 0.5^o\div1^o$ for
a Coma-like cluster. 
Taking as a reference the HESS sensitivity at $1 \; TeV$ for sources
uniformly extended on $\vartheta_{ext} \sim 1^o$, which is roughly
$\sim 3 \times 10^{-12} {\rm erg/cm^2/s}$ for 10 hours exposure, we
can estimate the on source exposure time needed for a detection
as a function of the parameter $\eta$:  
$$
t_{obs} \approx 50 \left(\frac{\eta}{0.1}\right)^{-2} \left(
\frac{\vartheta_{ext}}{0.5^o} \right) hrs. 
$$
Thus, a few tens of hours of exposure time seem adequate to detect a
Coma-like cluster in the most optimistic case in which $\eta \approx
0.1$, while deeper observations ($\gtrsim 100$ hours) are needed to
probe more conservative flux levels (e.g. lower values of $\eta$ 
and/or a steeper proton spectrum).  
 
The detection of TeV photons with a brightness profile correlated with
the gas density profile would hint to the presence of relativistic
hadrons in the ICM and would allow to finally measure the cluster non
thermal energy content.  

\subsection{Gamma rays from Inverse Compton Scattering (ICS)}

The detectability of the ICS gamma ray signal from 
electrons accelerated at either merger or accretion related shocks 
crucially depends on the strength of the shocks formed in either 
one of these cases. Only strong shocks can accelerate particles with
spectra hard enough to result in an appreciable gamma ray emission. 

The LogN-LogS diagram of clusters of galaxies as gamma ray sources was
calculated in \cite{noi3} using the PS based formalism described 
in Sec. \ref{sec:shocks}. The balance between injection and energy
losses drives electrons towards their time independent equilibrium
distribution, with a spectrum one power steeper than the injection
spectrum. 
The results of \cite{noi3} are shown in Fig. 5 (left panel), where the 
typical sensitivities of GLAST, AGILE and EGRET are also indicated. 
The calculations refer to the case of a constant efficiency of
electron acceleration of $5\%$, independent of the Mach number of the
shocks. It is worth stressing that the electrons accelerated according
with this recipe provide a negligible contribution to the diffuse 
radio emission and to the hard X-ray emission. Within these
assumptions, about $\sim 50$
clusters should be detected in gamma rays above 100 MeV by GLAST, 
equally distributed between merging and accreting clusters. AGILE 
on the other hand might be able to detect $\sim 10$ clusters.
Moreover, according to these predictions, no cluster should 
have been detected by EGRET, in agreement with what observed
\cite{olaf}.  
The self--consistent determination of the shock Mach numbers and of
the slope of electron spectra is the crucial ingredient of these
calculations \cite{noi2,noi3,berrington}. Previous approaches, relying 
on the incorrect assumption that all the structure formation shocks
are strong (and thus the electron spectrum is always the canonical
$E^{-2}$ power law), led to exceedingly optimistic predictions
on the number of detectable sources \cite{wl,totani}. 
 
The efficiency of acceleration of electrons at shocks is an unknown 
quantity. The value of 5\% which is generally adopted is a reasonable 
(maybe even optimistic) guess, but nothing more. Recent claims
\cite{keshet1} that this efficiency could be measured in the case of 
supernova shocks also appears rather optimistic, given the numerous
assumptions that such as estimate is based upon (for instance 
the Mach number of the shock is very different in the two cases, the 
shock modification could play a crucial role, the strength of the 
magnetic field could be dramatically different and for remnants
showing TeV emission it is hard to distinguish between leptonic and
hadronic scenarios). 

In Fig. 5 (right panel) we show the ICS gamma ray spectrum in the TeV
domain as expected from a Coma-like cluster of galaxies at 100 Mpc
distance. The effect of the gamma ray absorption in the infrared
background is illustrated by the difference between the solid lines
(with absorption) and dashed lines (without absorption).
From top to bottom, the lines refer to three different cases: 1) a merger
between two clusters with masses $10^{15}M_\odot$ and $10^{13}M_\odot$; 2)
an accreting cluster with mass $10^{15}M_\odot$ with a magnetic field
at the shock in the upstream region $0.1\mu G$; 3) an accreting cluster with 
mass $10^{15}M_\odot$ with a magnetic field at the shock in the upstream 
region $0.01\mu G$. 
The thick solid lines represent the sensitivities for a generic Cherenkov 
telescope array as calculated in \cite{HESSintent} for a point like
and a $\sim 1^o$ extended source, which is roughly the apparent size
of two merging clusters at 100 Mpc distance. Accreting clusters are 
even more extended ($\sim 2 - 3^o$) and thus harder to detect.  
More specifically, a telescope like HESS could probe the flux levels
considered here only after $> 50 - 100$ hours of observation. 

\subsection{Gamma rays from interactions of Ultra High Energy protons
  in the ICM} 
\label{sec:ultra}

Another channel for the production of gamma rays in the ICM is related
to the possibility that Ultra High Energy protons ($E>10^{18} eV$) 
may be present in the ICM \cite{felix,dario}. 
Before escaping the cluster, a proton with energy $E_p$ loses a
fraction of its energy by interacting with cosmic microwave background
photons, producing electron--positron pairs via the Bethe-Heitler
process. Pairs are produced with energy $\sim (2 m_e/m_p)
E_p$. These pairs rapidly cool via synchrotron emission and ICS, 
producing radiation in the hard--X and TeV gamma ray bands
respectively. The spectrum of the radiation is a power law with  
a hard spectrum (photon index $\Gamma \sim 1.5$), so that no
significant emission is expected in the GeV domain. 
Possible sources of very high energy protons include AGNs
\cite{felix} or cluster accretion shocks \cite{susumu}. The
corresponding gamma ray flux is however extremely uncertain and in the
absence of efficient confinement up to such high energies the flux may
be too low to be ever detectable. 

\section{Diffuse gamma ray background from structure formation}
\label{sec:back}

In the last decade clusters of galaxies have often been proposed as
the sources of the extragalactic diffuse gamma ray background, but more
accurate calculations have always lowered their contribution 
to a negligible fraction. This has been a hot topic in the 
field, with numerous papers addressing the issue
\cite{dar,bbp,lw,noi2,totani,keshet1,miniatifondo}. 
In this section we briefly summarize the relevant aspects of the
problem.  

\begin{figure}[thb]
 \begin{center}
  \mbox{\epsfig{file=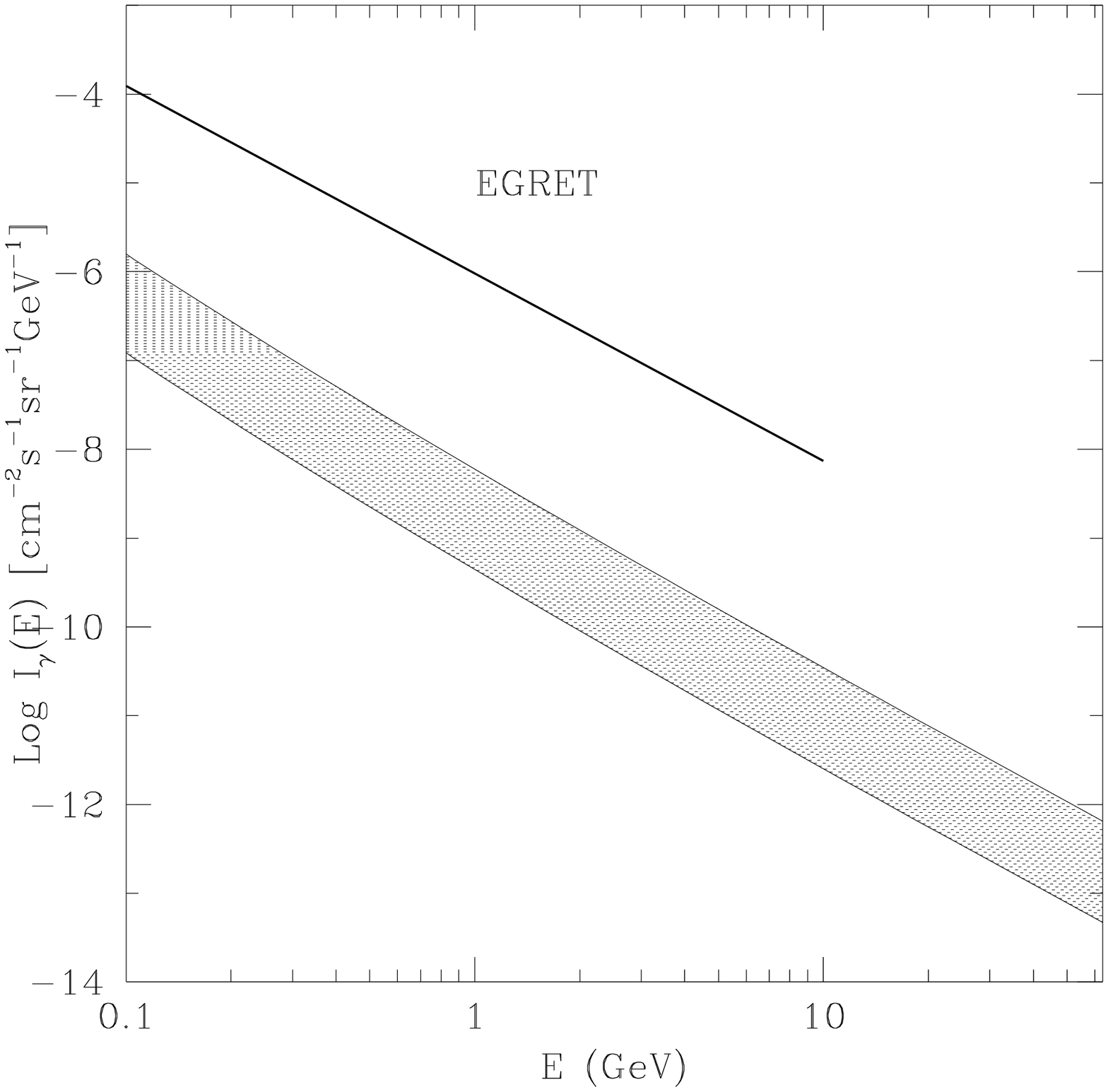,width=6.cm}}
  \hspace{0.3cm}
  \mbox{\epsfig{file=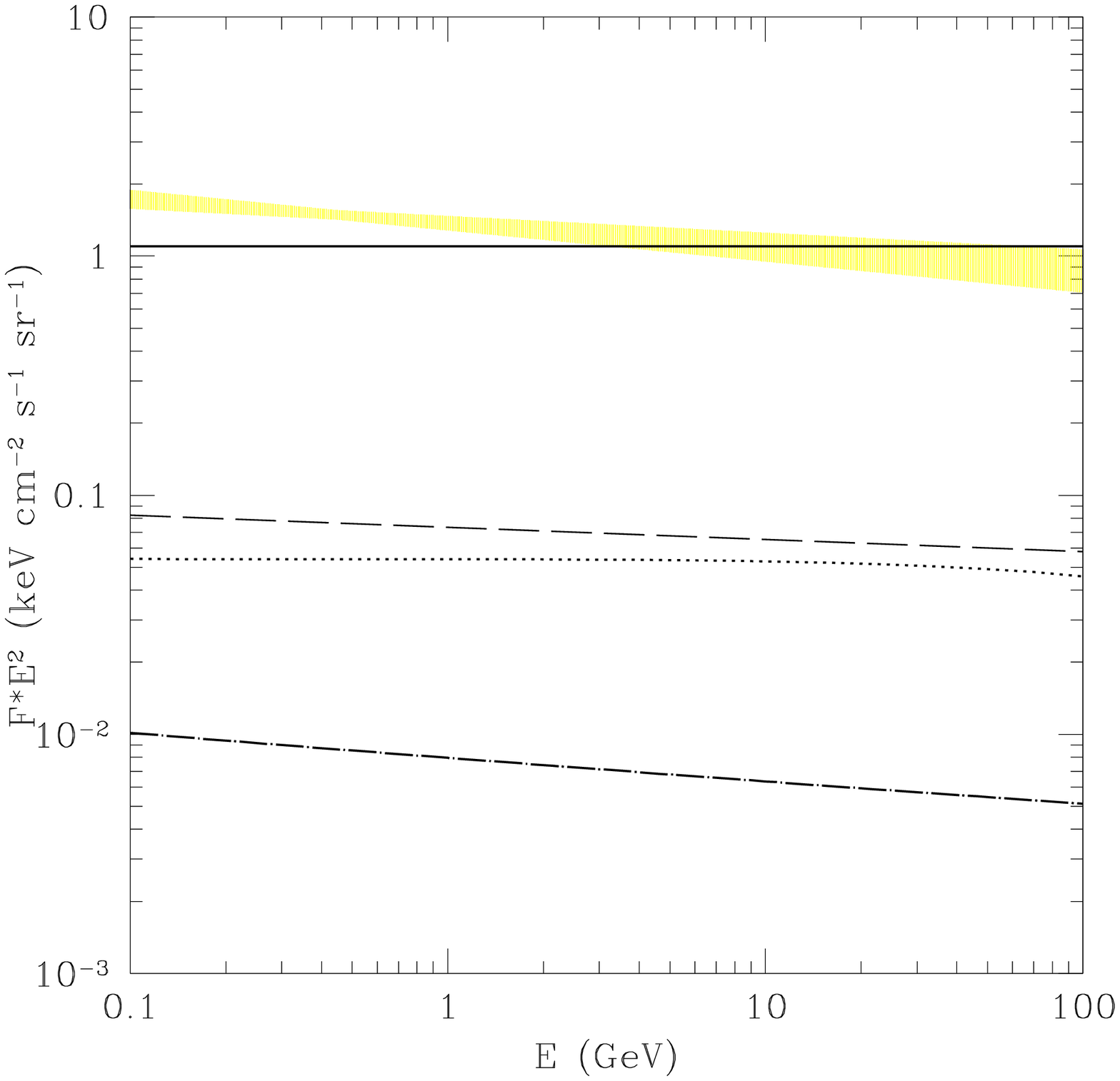,width=6.cm}}
  \caption{Diffuse gamma ray background due to clusters of
  galaxies. {\it Left Panel}: contribution of hadronic interactions
  $^{32}$
  with its uncertainty (shaded region). {\it Right Panel}:
  contribution from ICS of relativistic electrons. In both plots we
  also show the flux of diffuse extragalactic gamma ray background
  measured by EGRET $^{119}$.}
\label{fig:gamma}
 \end{center}
\end{figure}

Early interest in clusters of galaxies as potential sources of
the diffuse gamma ray background was stimulated by the discovery of
cosmic ray confinement in the intracluster volume
\cite{bbp,vab}, which increases the effective column density for
the process of pion production. In \cite{bbp} many potential sources
of cosmic rays were considered, and the contribution of 
clusters of galaxies to the diffuse background observed by EGRET
was found to be negligible. This result was later confirmed in
a more detailed calculation \cite{colablasi}, where however 
the spectrum was chosen {\it ad hoc} as a rather flat power
law ($E^{-\alpha}$ with $\alpha=2.2-2.4$), resembling the spectrum of
particles accelerated at astrophysical shocks and having in mind the 
secondary electron model for the origin of radio halos. 
The diffuse flux of gamma rays as calculated in \cite{colablasi} is
plotted in Fig. \ref{fig:gamma} (Left panel), where the shaded region
provides an estimate of the uncertainties arising from source
luminosities and cosmological models. In the light of the most recent
understanding of particle acceleration in clusters of galaxies, these
fluxes should be considered as upper limits to the actual hadronic
contribution of clusters of galaxies to the diffuse extragalactic
gamma ray background.

As discussed in the previous sections, electrons are also expected to
be accelerated in the ICM. Here we concentrate our attention on
electrons accelerated at either merger or accretion shocks. Electrons 
accelerated at each merger or accretion event lose energy through
ICS on the photons of the cosmic microwave background, which get
upscattered to gamma ray energies.

The diffuse flux of gamma radiation due to ICS of electrons
accelerated in mergers (dashed line) and during accretion (dotted
line) is plotted in Fig. \ref{fig:gamma} (right panel), where an
acceleration efficiency (for electrons) of $5\%$ has been 
assumed \cite{noi2}. The observed diffuse flux is represented by the 
shaded region \cite{sreekumar} (a recent re-analysis of the EGRET data
led to a slightly lower extragalactic diffuse gamma ray background
\cite{egretnew}). 
The calculation of the contribution of merger shocks requires to fix a
minimum mass of the merging clusters. The dashed line refers to
a minimum mass $\sim 10^{11} M_{\odot}$, while a more realistic result
is plotted as a dash-dotted line, where the minimum mass is fixed at
$10^{13}M_\odot$ (roughly the mass of a galaxy group).
In the same figure we plot for comparison the predictions of \cite{lw} 
(solid line), where the same acceleration efficiency was adopted. 
Several arguments have
been presented in \cite{noi2} to explain the difference between
their predictions, illustrated in Fig. \ref{fig:gamma} and those of
\cite{lw,totani}, the basic points being, as already discussed above,
that it is crucial to properly evaluate the Mach numbers of the merger
shocks and that it is somewhat important to keep track of the redshift
evolution of the merger rate.

The contribution of merger shocks to the diffuse extragalactic gamma
ray background is therefore likely to be in between the dashed and the
dash-dotted line.

The flux of gamma radiation from both mergers and accretion is a factor 
$\sim 10-100$ smaller than the observed background and smaller than the
flux predicted in \cite{lw,totani}, by the same factor. An
acceleration efficiency of the order of $50\%$ should be adopted in
order to reproduce observations. This would be unreasonable for
electrons as accelerated particles, and would violate our initial
assumption of shock acceleration in the linear regime (no backreaction  
of the accelerated particles on the shock). 

In \cite{keshet2}, a reevaluation of the diffuse
gamma ray background from large scale structure shocks was carried out 
by using N-body simulations. The results seem to be consistent with 
those plotted in Fig. \ref{fig:gamma}. In \cite{keshet2} the
authors emphasize the difficulties in the identification of shocks
with Mach number below $10$ (and the impossibility to detect shocks
with Mach numbers below $\sim 3-4$). 

\section{Conclusions}
\label{sec:concl}

Clusters of galaxies are sites where non-thermal activity has been
observed, mainly in the form of radio and hard X-ray radiation,
resulting from synchrotron and inverse Compton emission of
relativistic electrons respectively. The sources of such
electrons, the acceleration mechanisms involved in their production
and the coexistence of a hadronic component in the ICM are subjects 
of intense debate, which we summarized here. At the present time
observations seem to require a contribution from particle 
reacceleration through either resonant or non resonant particle-wave 
interactions. Although the general principle that this
acceleration process is based upon is well understood, several
details deserve more attention, especially in the direction of
achieving a fully self-consistent picture of particle acceleration
in the presence of plasma instabilities in the intracluster medium. 
The development of this line of investigation appears to be even
more justified by the uprising of modern observational facilities
such as LOFAR and LWA, operating at low radio frequencies, where
giant radio halos are expected to be more frequent
\cite{cassanoiosetti}.

Both primary and secondary models of radio halos run into
problems in explaining self--consistently morphological, spectral and 
statistical properties of extended radio halos. However, it should be
clear that most of these constraints derive from direct
analyses of the data on the Coma cluster and a few other
clusters that have been studied in a systematic way, and might reveal 
themselves as incomplete when larger samples will become available. 

The detection of gamma rays from nearby clusters of galaxies
would provide us with a powerful tool to discriminate among 
the various models proposed to explain observations. 
An overview of the existing predictions and models appears quite timely, 
since several gamma ray telescopes are starting to operate 
(HESS, MAGIC, VERITAS) and some others are coming soon (GLAST, AGILE).     

The basic mechanisms for the production of gamma radiation in the ICM
are 1) the decay of neutral pions produced in inelastic $pp$ collisions
and 2) ICS of high energy electrons on the photons of the
cosmic microwave background.

Both electrons and protons (or more in general nuclei) are expected to
be accelerated at the shock fronts that form during the process of
structure formation. These shocks are typically classified as merger
shocks and accretion shocks, depending on whether they are generated
in the inner regions of the parent clusters (which makes them weaker
due to the high temperature of the ICM) or in the outskirts and 
filaments, where fast, cold, non virialized plasma falls into the 
gravitational potential well of the parent cluster. A small
fraction of merger shocks may also have relatively large Mach
number.

The contribution of the hadronic channel to gamma ray emission
is rather model dependent 
because there may be several classes of sources that may contribute
at different levels and with different spectra of accelerated
particles. If acceleration takes place at the shocks that develop 
during structure formation, then an estimate of the energy in the 
form of non-thermal particles may be obtained from the merger history
of a cluster as described by the hierarchical model of structure
formation. At each merger
event that a cluster suffers a new population of hadrons is injected but 
the pre-existing cosmic ray population, diffusively confined in the 
ICM \cite{bbp,vab}, is also reaccelerated \cite{noi1}. This 
combination of effects was shown to result in non-power law spectra.
Due to the weakness of shocks in the inner parts of galaxy clusters, 
the spectra of accelerated particles are typically steep, and the
proton spectrum accumulated in the ICM as a result of mergers does not
lead to intense gamma ray emission \cite{noi1,berrington}. 
Acceleration at accretion 
shocks in the outskirts of clusters is more promising 
(the Mach number is $\sim 10-100$) and the particles could be advected
into the cluster and trapped there for cosmological times. 
The gamma ray emission due to $\pi^0$ decay in this scenario might 
be detected by GLAST, depending on the acceleration efficiency at the
accretion shock. If additional hadrons are injected by sources
in the ICM (galaxies, AGNs) the gamma ray flux due to hadronic
interactions may be somewhat higher \cite{vab,colablasi,ensslin97}. 
Rather strong limits on the energy content allowed to be in the form 
of non-thermal hadrons can be imposed (mostly adopting the Coma 
cluster as the prototype cluster) by using the EGRET upper limits
\cite{colablasi,blasi99,pasqualemerger,pfrommer}, 
upper limits on the flux of TeV photons \cite{blasi99} and the
existence of the cutoff in the radio spectrum at
GHz frequency \cite{reimers}. Typically these methods lead to
conclude that not more than $\sim 10-30\%$ of the energy of the
cluster can be in the form of a non-thermal population of hadrons. 
These limits are however rather model dependent and should be used
with care. If the amount of hadrons is not much smaller than this
value, and particle spectra are hard enough ($\propto E^{-2}$),
forthcoming observations in the TeV domain might finally detect the
first gamma rays from clusters. 

The detection of gamma rays generated through ICS of high
energy electrons accelerated at merger and accretion shocks appears
somewhat more promising. As illustrated in Fig. \ref{fig:logNlogS},
$\sim 10-50$ clusters should be detectable with GLAST in the energy
range $>100$ MeV. This prediction is consistent with the fact that
EGRET has not detected any cluster \cite{noi3,olaf}. The detection of
gamma rays in the TeV energy range appears more problematic, due to
the extended size of the emission region. 

The contribution of clusters of galaxies to the extragalactic diffuse 
gamma ray background is less than $1-10\%$ of the flux measured in
\cite{sreekumar}. As better observations will become available, it is
possible that the classes of sources that contribute to the
extragalactic gamma ray background may become better identified,
either by subtraction of known sources, or by identification of new
objects with a given gamma ray emission.

From the theoretical point of view the most important development in
the understanding of the non-thermal aspects of structure formation
will most likely come from the introduction of non linear effects in
diffusive particle acceleration at shocks. These effects are showing
all their importance for the case of supernova remnants as cosmic ray
accelerators and it is likely that a similar role will arise 
in the future for large scale structure shocks.

\section*{Acknowledgements}

The authors are very grateful to F. Aharonian, T. Jones, O. Reimer and
E. Resconi for constructive comments on the manuscript. 
SG gratefully acknowledges support from the Alexander von Humboldt
foundation. The research work of PB and GB was partially funded through
grant PRIN2004. GB acknowledges partial support from grants
PRIN2005 and PRIN-INAF2005.

\end{document}